\documentclass{article}
\usepackage{arxiv}

\usepackage[utf8]{inputenc} 
\usepackage[T1]{fontenc}    
\usepackage[hidelinks]{hyperref}       
\usepackage{url}            
\usepackage{booktabs}       
\usepackage{amsfonts}       
\usepackage{nicefrac}       
\usepackage{microtype}      
\usepackage{float}
\usepackage{doi}
\usepackage{subfigure}
\usepackage{multirow}
\usepackage{amsmath}
\usepackage{graphicx}
\usepackage{setspace}
\usepackage{amssymb} 
\usepackage{xcolor} 
\usepackage{subcaption}
\usepackage{colortbl}

\usepackage[mathlines]{lineno}
\usepackage{appendix}
\usepackage{makecell} 

\usepackage{xspace}
\usepackage{diagbox}
\usepackage{adjustbox}
\usepackage{dirtytalk}
\usepackage{dsfont}
\usepackage{arydshln}
\usepackage{enumitem}

\definecolor{lavender}{rgb}{0.9, 0.9, 0.98}

\newcommand{\GC}{genderComputer\xspace}

\newcommand{\monotf}{\texttt{MonoT5}\xspace}
\newcommand{\rankvic}{\texttt{RankVicuna}\xspace}
\newcommand{\rankzep}{\texttt{RankZephyr}\xspace}
\newcommand{\mistral}{\texttt{Mistral-7B}\xspace}
\newcommand{\llama}{\texttt{LLaMA-3-8B}\xspace}
\newcommand{\accdate}{\texttt{AccDate}\xspace}
\newcommand{\accday}{\texttt{Acc1Day}\xspace}

\makeatletter
\def\adl@drawiv#1#2#3{%
        \hskip.5\tabcolsep
        \xleaders#3{#2.5\@tempdimb #1{1}#2.5\@tempdimb}%
                #2\z@ plus1fil minus1fil\relax
        \hskip.5\tabcolsep}
\newcommand{\cdashlinelr}[1]{%
  \noalign{\vskip\aboverulesep
           \global\let\@dashdrawstore\adl@draw
           \global\let\adl@draw\adl@drawiv}
  \cdashline{#1}
  \noalign{\global\let\adl@draw\@dashdrawstore
           \vskip\belowrulesep}}
\makeatother

\makeatletter
\g@addto@macro\UrlBreaks{%
  \do\/\do\-\do\_\do\?\do\&\do\=\do\#%
}
\makeatother

\title{Gender Disparities in StackOverflow's Community-Based Question Answering: A Matter of Quantity versus Quality}

\author{
Maddalena Amendola\textsuperscript{1},
Cosimo Rulli\textsuperscript{2},
Carlos Castillo\textsuperscript{3,4},
Andrea Passarella\textsuperscript{1},
Raffaele Perego\textsuperscript{2} \\
\\
\textsuperscript{1}IIT-CNR, Pisa, Italy \\
\textsuperscript{2}ISTI-CNR, Pisa, Italy \\
\textsuperscript{3}Universitat Pompeu Fabra, Barcelona, Spain\\
\textsuperscript{4}ICREA, Barcelona, Spain
}

\begin{document}

\maketitle

\begin{abstract}
Community Question-Answering platforms, such as \textsc{Stack Overflow} (SO), are valuable knowledge exchange and problem-solving resources. These platforms incorporate mechanisms to assess the quality of answers and participants' expertise,  ideally free from discriminatory biases. However, prior research has highlighted persistent gender biases, raising concerns about the inclusivity and fairness of these systems. Addressing such biases is crucial for fostering equitable online communities.
While previous studies focus on detecting gender bias by comparing male and female user characteristics, they often overlook the interaction between genders, inherent answer quality, and the selection of ``best answers'' by question askers.
In this study, we investigate whether answer quality is influenced by gender using a combination of human evaluations and automated assessments powered by Large Language Models.
Our findings reveal no significant gender differences in answer quality, nor any substantial influence of gender bias on the selection of ``best answers." 
Instead, we find that the significant gender disparities in SO's reputation scores are primarily attributable to differences in users' activity levels, e.g., the number of questions and answers they write.
Our results have important implications for the design of scoring systems in community question-answering platforms. In particular, reputation systems that heavily emphasize activity volume risk amplifying gender disparities that do not reflect actual differences in answer quality, calling for more equitable design strategies.
\end{abstract}

\section{Introduction}
\label{sec:intro}

Community Question\&Answering (CQA) platforms have become invaluable sources of knowledge and skill-sharing, evolving from platforms for exchanging information to tools for showcasing expertise.
One well-known example is \textsc{Stack Overflow}\footnote{\url{https://stackoverflow.com/}} (SO), the most extensive programming community, which is not only used by developers to find answers to questions but is also used by companies to identify and recruit experts in specialized fields of computer science~\cite{maftouni2022thank}.
Recent studies have highlighted several challenges within \textsc{Stack Overflow}, particularly concerning inclusivity.
The platform is perceived as unwelcoming to minority groups, especially women, who make up only about 10\% of the user base \cite{may2019gender,dev2019quantifying,maftouni2022thank,scheltens2022representation}.
Aggravating the problem, \textsc{Stack Overflow}'s reputation system (a core metric used to evaluate and showcase user expertise) favors male users, who, on average, achieve SO-Reputation scores twice as high as their female counterparts \cite{wang2018understanding,may2019gender}.
While some studies attribute this disparity to men's higher activity levels, others suggest that the platform's predominantly male user base may contribute to biased voting and rating patterns \cite{may2019gender,wang2018understanding}. 
As a result, women may perceive \textsc{Stack Overflow}  as a hostile environment that discourages them from actively participating and instead encourages them to adopt strategies such as preferring to engage with posts where other women are present \cite{morgan2017programming,ford2017someone}.
Online platforms that embrace and do not hinder the participation of minority groups, especially women, are necessary to foster inclusive online spaces and increase the diversity and quality of the content.

This study explores the interplay between answer quality, the selection of best answers (i.e., those marked as ``accepted'' by askers), and \emph{perceived gender} within the \textsc{Stack Overflow} community.
It aims to identify whether the recognition of contributions and the selection of best answers is driven solely by content quality or also influenced by the perceived gender of the contributor. By addressing these questions, this work contributes to understanding bias in \textsc{Stack Overflow} and highlights ways in which such platforms might promote greater inclusivity.

Because \textsc{Stack Overflow} does not collect or display gender identities of platform users, we estimate perceived gender based on usernames and country data using \GC\footnote{\url{https://github.com/tue-mdse/genderComputer}} \cite{vasilescu2012gender}, a widely adopted tool in gender bias studies.
This approach, which we acknowledge has the limitation of assuming a binary gender, seems to approximate the kind of gender inference that users themselves might implicitly perform when viewing other contributors' usernames.
In other words, our focus is not on actual gender identity, which users do not declare on this platform, but instead on exploring if and how gender \emph{perceptions} influence community behavior.

Additionally, given their strong reasoning and coding capabilities, we leverage state-of-the-art Large Language Models (LLMs) to assess the quality of male and female answers concerning \textsc{Stack Overflow} questions.

We first perform two human assessments to validate the methodology intended to approximate human judgment: (i) the accuracy of gender inference by \GC, and (ii) the assessment of answer quality for specific questions provided by male and female users.
The human evaluation supports \GC’s effectiveness in inferring user gender, achieving a precision of approximately $90\%$ for men and $80\%$ for women. Furthermore, by comparing the selection of the ``best answer'' by \textsc{Stack Overflow} users and LLMs to human assessments, we observe that state-of-the-art LLMs achieve alignment rates of up to $76\%$, indicating good reliability in assessing answer quality.

Next, we conduct a statistical analysis on relevant aggregated features (i.e., SO-Reputation, views, and downvotes) to compare men's and women's behaviors.
This analysis reveals that men generally have higher SO-Reputation and higher average answer scores, aligning with their greater overall activity.
On the other hand, when we restrict our analysis to questions with both male and female responses, women show higher activity levels, supporting the hypothesis of homophilic behavior.

Finally, we compare LLM-generated evaluations with those from \textsc{Stack Overflow}, assessing how often a male answer is accepted over a female one that the LLM selects as the best answer, and vice versa.
We find that (i) LLMs agree with \textsc{Stack Overflow}’s accepted answers between $47\%$ and $68\%$ of the time; and (ii) the frequency of choosing a male-authored answer over a female-authored one, or vice versa, shows no systematic pattern, with differences in selection rates remaining below $2\%$. This result highlights that the asker's subjective preferences may also influence discrepancies in selecting the best answer.

The overall analysis reveals that male and female users provide answers of equal quality, with disparities in recognition metrics primarily linked to the different activity levels and the design of the reputation system, which heavily emphasizes activity volume. 

The main contributions of our study are the following:
\begin{itemize}
\item \textbf{Analysis of Features-Based Gender Differences:} our statistical analysis supports previously observed gender disparities, including male dominance in key features, and prior findings on women's homophilic behavior.  
\item \textbf{LLM-Based Evaluation of Answer Quality and Selection:} leveraging LLMs, we analyze answer quality differences across genders and investigate potential gender-related biases in the selection of the ``best answer''.  

\end{itemize}

To our knowledge, this is the first study to directly compare the quality and correctness of male and female answers, while previous research primarily suggested bias based only on a statistical analysis of user features, neglecting the quality of answers. 
The paper is structured as follows: Section~\ref{sec:relatedwork} reviews previous research on biases in \textsc{Stack Overflow}. Section~\ref{sec:datasets} introduces the \textsc{Stack Overflow} datasets, details the gender inference methodology using \GC, and describes dataset partitions. Section~\ref{sec:methodology} outlines our methodology, including Features-based (Section~\ref{subsec:features_analysis}) and LLM-based analyses (Section~\ref{subsec:llm_analysis}). Section~\ref{sec:human} presents human evaluations conducted via Amazon Mechanical Turk\footnote{\url{https://www.mturk.com/}}. Section~\ref{sec:results} discusses the results, while Section~\ref{sec:conclusion} concludes with key findings and implications.

\section{Related Work}
\label{sec:relatedwork}

CQA platforms, such as \textsc{Stack Overflow}, are designed to promote knowledge sharing and user engagement. These platforms are often perceived as meritocratic and free of gender barriers due to their openness and transparency \cite{wang2018understanding}. 
However, research indicates that \textsc{Stack Overflow} tends to be a male-dominated environment \cite{wang2018understanding}. 
Ford et al. \cite{ford2016paradise} conducted semi-structured interviews and surveys, uncovering that women face several more barriers than men. These barriers include doubts about their expertise and feeling overwhelmed by the competitive environment. Jay Hanlon, vice president of community growth at \textsc{Stack Overflow}, acknowledged the presence of race and gender biases, stating that many perceive \textsc{Stack Overflow} as hostile or elitist, particularly newer coders, women, people of color, and other marginalized groups \cite{hanlon2018stack}.
\textsc{Stack Overflow} has frequently been criticized for being a harsh and unfriendly environment \cite{maftouni2022thank}. 

\textsc{Stack Overflow} included analysis on gender in its Developer Survey\footnote{\url{https://survey.stackoverflow.co/}} only between 2015 and 2022. During this period, the share of women respondents remained stable at approximately 10\%. This persistent imbalance has been a recurring theme across the surveys. For instance, the 2019 survey confirmed that all developer categories had dramatically more men than women, in line with broader research showing that women leave tech jobs at higher rates than men. Moreover, \textsc{Stack Overflow} asked nearly 80,000 users what aspects of the platform they would most like to change, revealing gender-based differences in perceptions: men associated the platform with terms like ``officia'', ``complex'', and ``algorithm'', while women described it as ``condescending,'' ``rude,'' and ``assholes''~\cite{brooke2019condescending}. The 2020 survey further underscored the importance of inclusion and retention.
When asked about desired changes to \textsc{Stack Overflow} itself, women were more likely than men to highlight the need for better communication norms. Finally, the 2022 survey reported that women were less likely to consider themselves part of the \textsc{Stack Overflow} community, illustrating how the platform’s perceived inclusiveness remains a concern. Since 2023, questions about gender have been removed from the survey due to privacy concerns. Anecdotal reports also portray \textsc{Stack Overflow} as unwelcoming, with frequent criticisms of reputation-based elitism and harsh communication practices.\footnote{\url{https://meta.stackoverflow.com/questions/366665/does-stack-exchange-really-want-to-conflate-newbies-with-women-people-of-color}}

Common challenges women face include fear of negative feedback, lower confidence in their programming skills, an unwelcoming environment, inappropriate language, the competitive nature of the platform, and lack of peer support \cite{scheltens2022representation}.
Furthermore, evidence suggests that men benefit more from the current reputation system, which is biased against women. This disparity arises from gender differences in participation: women are more likely to ask questions, while men are more likely to provide answers and cast votes. The system favors answering questions, disadvantaging women due to their higher tendency to ask questions \cite{wang2018understanding, may2019gender}. Studies show that the average woman has roughly half the SO-Reputation points of the average man \cite{may2019gender, brooke2021trouble}. In fact, male users tend to receive higher scores for their answers, suggesting that biases in scoring contribute to gender disparities in SO-Reputation \cite{brooke2021trouble}.
Additionally, votes on questions and answers are influenced by SO-Reputation bias, where users are more likely to vote positively on content from users with higher SO-Reputations, regardless of content quality \cite{dev2019quantifying}.
Finally, Ford et al. \cite{ford2017someone} suggest that the presence of more women in a thread creates a supportive environment, boosting female participation and fostering a sense of belonging. Women become more active after engaging in peer parity posts, defined as interactions where individuals can identify with at least one other peer \cite{morgan2017programming}. Brooke \cite{brooke2021trouble} concludes that \textsc{Stack Overflow} users tend to interact with others of the same or similar gender, indicating a gender-based organization in user interactions. 

In conclusion, while \textsc{Stack Overflow} aims to be an open and meritocratic platform, significant gender biases persist, affecting user experiences and outcomes. Addressing these biases is crucial to creating more inclusive and supportive communities.
Despite increased research efforts over the past decade, women’s representation on \textsc{Stack Overflow} remains low~\cite{scheltens2022representation}. Proposed solutions include adjusting the reputation system to value question contributions as highly as answers~\cite{wang2018understanding}. Such a redesign could reduce the gender gap in SO-Reputation scores by nearly half~\cite{may2019gender}, promoting a more balanced and inclusive experience.

To our knowledge, this is the first study that focuses explicitly on the quality and correctness of the answers provided by male and female \textsc{Stack Overflow} users. By leveraging human-, feature-, and LLM-based evaluations, we show that (i) there are no quality differences in answers provided by male and female users; (ii) askers' selection of the best answer is not influenced by gender bias but rather by subjective and objective factors; and (iii) recognition disparities stem from differences in participation patterns and a reputation system that heavily favors high activity levels.
\section{Data Collection and Annotation}
\label{sec:datasets}
In this section, we describe the structure of the \textsc{Stack Overflow} community, outline the methods for gender inference, and explain the process of creating curated datasets tailored to our analysis.

\subsection{The Stack Overflow community}
\textsc{Stack Overflow}, the most extensive and widely-used programming community, is the largest community in the Stack Exchange\footnote{\url{https://stackexchange.com/}} network, a group of CQA portals that provide quarterly data dumps\footnote{\url{https://archive.org/details/stackexchange}} of community activity, covering all content from the creation of each site to the present. 

\textsc{Stack Overflow} registered its first question in 2008 and hosts over 24 million questions, 36 million answers, and 26 million users. Users on \textsc{Stack Overflow} post questions by specifying a title, a detailed body, and a set of tags chosen from a predefined list, which categorize the question's topic. By using tags, community members can quickly find open questions where they can offer help in solving problems. As technology evolves rapidly, the distribution of tag usage follows a power-law distribution (c.f. Figure 2 in~\cite{chen2019stack}): a few tags are used frequently, representing the major topics discussed on the platform, while the majority are used infrequently, typically reflecting more specific or niche problems. 
Finally, the community can upvote or downvote questions and answers, and importantly, only one answer can be marked as \emph{accepted}, thus identifying the \emph{best solution} for the asker. As users receive more upvotes and have their answers accepted, their SO-Reputation score increases, reflecting their expertise and ability to provide high-quality solutions.

A comprehensive overview of the features available in the \textsc{Stack Overflow} dataset is presented in the Appendix.

\subsection{Gender Inference}
\textsc{Stack Overflow} does not require users to specify their gender, leading researchers interested in gender-based analysis to develop tools and methodologies to infer this information.
The \GC  tool is designed to infer a user's gender (i.e., male, female, or unknown) based on their name and country \cite{vasilescu2012gender}.
It relies on statistics about the occurrence of names in specific countries and can even handle modified usernames, such as ``w35l3y''.
We adopt \GC in our work as it is widely used in research studies focusing on gender biases in \textsc{Stack Overflow}~\cite{wang2018understanding,scheltens2022representation,may2019gender,brooke2021trouble,ford2017someone}.

Our aim is not to classify users’ actual gender identities, but rather to approximate the perceived gender that a typical platform user (e.g., an asker) might infer based on personal information like name and country. In real-world settings, this perception is likely to be binary and based on cultural assumptions. Thus, using binary gender labels provides a pragmatic proxy for understanding how biases may manifest in a community where such perceptions influence behavior.

However, we acknowledge several ethical implications and limitations when using a tool that infers (binary) gender based on names and country-specific statistics.
Firstly, the tool assumes that usernames are based on real names and that users disclose their actual location.
Some individuals, particularly women, may use pseudonyms or adopt male-sounding usernames to hide their gender and avoid bias or stereotypes~\cite{bruckman1996gender,szell2013women}.
Moreover, names can have different gender associations across cultures or linguistic groups within a country. A tool relying on name statistics from specific countries may misinterpret names in multicultural contexts, leading to incorrect inferences. 
As gender identification often depends on nuanced, contextual aspects, we employ a human evaluation process in Section~\ref{subsec:mt_gender} to evaluate \GC’s accuracy.
While binary simplification excludes non-binary and fluid identities, it aligns with our focus on measuring potential bias in the gendered perception of contributors.
The goal of our research is to promote inclusiveness in these platforms. Recognizing their mechanisms and limitations can help understand how biases against women can be addressed, which is a step in that direction.

Hereafter, we omit the adjective ``perceived'' before gender for brevity, though we always refer to inferred gender as the one most likely perceived by the community.

\subsection{Data sampling}

Given LLMs' complexity and resource-intensive nature, we streamline the dataset by considering only the questions and the answers posted from 2017 to the present. 
Moreover, we perform several pre-processing steps.

First, we use \GC to infer the gender of all platform users and discard those whose gender cannot be inferred. The entire dataset consists of 22 million users, but only 3 million contributed as answerers. From this subset, we excluded users without a recorded name or location, resulting in approximately 1 million users. Using \GC, we inferred the gender of 622,995 users, about 11\% of whom were female.
We consider the answers from the users selected in the previous step and the corresponding questions. At this point, we have all questions with at least one user for which we may infer the gender. 

To determine whether individuals exhibit bias when selecting accepted answers, we consider all questions with at least one male and one female answerer in the answer thread.
Specifically, we refine the answer set to include only those responses the asker might have seen at the time of acceptance.
This requires establishing a time threshold to decide which responses should be considered relevant for a given question, excluding responses posted as much as a year later. 
This process results in the creation of two distinct datasets:
\begin{itemize}
    \item \texttt{AccDate}: in the absence of exact acceptance time, we assume that users select the best answer when it is posted, based on the assumption that they generally seek timely solutions, actively monitor responses to their questions, and may receive notifications when an answer is posted.
    \item \texttt{Acc1Day}: we relax the constraints of the \accdate dataset by including all answers posted within one day of the accepted answer, allowing for a broader analysis of responses received in close temporal proximity.
\end{itemize}

\begin{table}[]
\centering
\caption{Number of questions, answers, users, and percentage of female (\%F) users across datasets.}
\label{tab:dataset_statistics}
\begin{tabular}{@{}lrrrr@{}}
\toprule
            & Questions & Answers  & Users  &  \%F \\ \midrule
\texttt{2008-2024}   & 10,640,406  & 14,176,892 & 622,995 & 10.97   \\
\texttt{2017-2024}   & 4,255,489   & 5,095,104  & 418,157 & 11.00     \\
\accdate     & 24,496     & 53,571    & 23,230  & 32.12     \\
\accday & 48,268     & 111,214   & 37,436  & 29.25   \\ \bottomrule
\end{tabular}
\end{table}

Table \ref{tab:dataset_statistics} presents statistics for the entire period alongside the two final datasets used in our analysis. 
Considering the entire community period \texttt{2008-2024}, female users (\%F column) account for approximately 11\% of answerers, consistently with previous findings \cite{may2019gender,dev2019quantifying,maftouni2022thank,scheltens2022representation}.
Focusing on a more recent period (\textit{2017-2024}) halves the number of questions and answers while maintaining a consistent percentage of female users. 
Limiting the dataset to questions with both male and female respondents before the accepted answer (\accdate) retains only 0.58\% of questions; extending the window by one day (\accday) increases this to 1.13\%.
Notably, in both datasets, the proportion of female respondents rises to approximately 30\%.
The same trend is observed when selecting questions with both male and female respondents over the entire period, potentially reflecting prior findings of homophilic behavior among women (i.e., a greater likelihood of responding when other female users have already participated), thereby increasing their presence in these threads~\cite{morgan2017programming,ford2017someone}. We further investigate this in Section~\ref{subsec:features_analysis_res}.

Given the large scale of the \textsc{Stack Overflow} dataset and the resulting computational complexity, we restrict our comparison of the \accdate and \accday subsamples to the \texttt{2017-2024} dataset, from which both are derived. 
Table~\ref{tab:topic_coverage} presents the topic coverage of the selected datasets across different percentiles (Pct column), based on tag occurrences in the full \texttt{2017-2024} dataset. At the 50\textsuperscript{th} percentile, all tags with at least 9 occurrences are included, representing a broad set of tags with relatively low frequency. For lower percentile values, topic coverage is limited. However, as previously highlighted, the field of computer science evolves rapidly, resulting in an exponential increase in tags with shallow frequency. In analyzing topic coverage, we aim to ensure that the main topics, identified by tags with very high occurrences, are represented. Substantial coverage starts from the 90\textsuperscript{th} percentile, which includes nearly all of the primary topics discussed on the platform.

\begin{table}[]
\centering
\caption{Topic coverage of the \accdate and \accday datasets relative to the \texttt{2017-2024} dataset across different percentile thresholds.}
\label{tab:topic_coverage}
\begin{tabular}{rrr@{\hspace{2.5\tabcolsep}}rr}
\toprule

\multirow{2}{*}{Pct} & \multirow{2}{*}{Min. Occ.} & \multirow{2}{*}{\#Tags} & \multicolumn{2}{c}{Coverage (\%)} \\ 
\cmidrule{4-5}
 & & & \accdate & \accday \\ 
\midrule
50.00      & 9      & 26,772   & 22.98   & 29.22     \\
75.00      & 37     & 13,224   & 41.13   & 50.69     \\
90.00      & 164    & 5,269    & 71.55   & 80.98     \\
95.00      & 424    & 2,630    & 89.54   & 94.33     \\
97.50      & 1,019  & 1,315    & 97.49   & 99.16     \\
99.00      & 2,697  & 528      & 99.81   & 100.00     \\ \bottomrule
\end{tabular}
\end{table}

\section{Methodology}
\label{sec:methodology}

To assess gender disparities on \textsc{Stack Overflow}, we adopt a twofold approach: (i) we first perform a \emph{Features-Based Analysis} to statistically compare user features, identifying patterns and trends that may indicate disparities; (ii) we then analyze gender differences by evaluating the quality of answers provided by male and female users, leveraging state-of-the-art LLMs. In this section, we outline the methodology adopted for both analyses.

\subsection{Features-based Analysis}
\label{subsec:features_analysis}

Previous research \cite{brooke2019condescending,brooke2021trouble,ford2016paradise,wang2018understanding} has highlighted the presence of gender biases on \textsc{Stack Overflow}, specifically against women, mainly by analyzing recognition metrics on the platform. 
Building on this body of work, we aim to statistically compare the distributions of key user-level features across genders and also assess how dataset filtering impacts the composition and characteristics of the resulting subsamples.

We consider the following features computed for each user over the entire available time span \texttt{2008-2024}: \emph{Reputation}, profile \emph{Views}, number of \emph{UpVotes} and \emph{DownVotes} cast, number of \emph{Answers} and \emph{Accepted Answers}, average answer \emph{Score}, and \emph{Average Delay}, defined as the mean response time in hours. This allows us to contextualize the activity, visibility, and recognition of users included in each sample relative to the broader platform population.

To statistically assess the differences, we apply the \emph{Mann-Whitney U} test, a non-parametric test used to evaluate whether two independent samples come from the same distribution. The test is conducted using commonly applied significance levels of $\alpha$: $0.05$ and $0.01$. When the p-value is lower than $\alpha$, the test rejects the null hypothesis, suggesting that the distributions of male and female users belong to different populations. Given the large user set and the significant dominance of male users, we enhance the test accuracy by performing 5,000 permutations and the Bonferroni correction \cite{dunn1961multiple}.

Finally, we analyze homophily in the answerer-asker network by measuring (i) gender assortativity~\cite{newman2002assortative} with the Spearman rank correlation, and (ii) the likelihood of users responding to questions from askers of the same gender, while accounting for the overall gender imbalance.

\subsection{LLM-based Analysis of Answers Quality}
\label{subsec:llm_analysis}

In this section, we evaluate the quality of answers provided by male and female users on \textsc{Stack Overflow}. By analyzing the proportion of cases where male answers were accepted over female answers and vice versa, we aim to uncover gender-based disparities in accepted answers.

Evaluating answer quality is a daunting task, as \textsc{Stack Overflow} questions span diverse domains
and a complete human assessment is impractical due to the large dataset. 
Instead, we use LLMs to compare the quality of multiple answers to user questions.

LLMs have demonstrated proficiency in tasks closely related to humans, namely ranking and relevance judgment. In the context of ranking, models based on Instructed Large Language Models (ILLMs) have shown effectiveness in ordering answers by their relevance to a given query and are acknowledged as the state-of-the-art solution. 
Regarding relevance judgment (i.e., the task of assessing the relevance of an answer given a user query) LLMs have proven to be highly effective~\cite{faggioli2023perspectives}, sometimes even providing more consistent judgments than humans.

Hence, we frame the task of selecting the best answers among those provided for a certain question as a ranking problem, where LLMs evaluate and order answers based on their relevance and quality concerning each question. This ranking allows us to identify the top answer for each question and compare it to the community’s accepted answer, providing a basis for assessing potential biases in selection.
We choose to employ two different kinds of LLM-based models, namely \textit{Re-ranker} models and \textit{ILLMs}.

\smallskip
\noindent \textbf{Re-Rankers}. The role of a re-ranker is to enhance the quality of a ranked list, typically by refining the selection made by the retriever or a preceding simpler ranking function \cite{pradeep2023rankzephyr}.  We use three state-of-the-art re-ranker models:
\begin{itemize}
    \item \monotf~\cite{nogueira2020document}: It adapts the T5 model~\cite{raffel2020exploring} to the problem of predicting query-document relevance scores.
    
    \item \rankvic~\cite{pradeep2023rankvicuna}: The first open-source LLM for listwise reranking in zero-shot scenarios.
    
    \item \rankzep~\cite{pradeep2023rankzephyr}: A state-of-the-art, open-source LLM for listwise zero-shot reranking that, in some cases, surpasses proprietary models such as RankGPT4. 
\end{itemize}

\smallskip
\noindent \textbf{Instructed Large Language Models}.
We employ two state-of-the-art open-source ILLMs, namely:
\begin{itemize}
    \item \llama\footnote{\url{https://huggingface.co/meta-llama/Llama-3.1-8B-Instruct}} \cite{dubey2024llama}:  it achieves state-of-the-art performance across various benchmarks, often comparable to leading models like GPT-4.
    \item \mistral\footnote{\url{https://huggingface.co/mistralai/Mistral-7B-Instruct-v0.1}} \cite{jiang2023mistral}: focused on efficiency, it outperforms larger models like LLaMa 2 13B.
\end{itemize}

To minimize potential biases from answer order \cite{liu2024lost}, we shuffle the answers for each question and create three distinct permutations of the set of answers. Each model generates a ranking for each permutation. Importantly, models receive only the question content (title and body), and answer text with no user information to avoid possible bias propagation through the model evaluation.

ILLMs can produce malformed answers that do not attain the provided instructions~\cite{pradeep2023rankvicuna} (e.g., returning different sets of answers or providing only a textual description). 
To ensure a fair comparison, we (i) set a low temperature of $0.1$ to promote deterministic outputs, and verified on a small subsample that indeed repeated runs consistently yielded identical answer rankings, and (ii) discarded all responses that did not follow the prompt instructions. Across both datasets, a portion of the questions was discarded during preprocessing due to inconsistent or incomplete model outputs.
After generating rankings across permutations, we aggregate results for each model individually. To keep the probability of agreement by chance below 5\%, we retained questions with fewer than five answers only if the same answer was ranked first in all three permutations, and questions with five or more answers only if the same answer appeared in the top position in at least two out of three permutations.

\section{Human Evaluation}
\label{sec:human}
We conduct a human evaluation to carefully assess two pivotal aspects of our study: (i) the accuracy of the \GC tool in inferring users' genders, and (ii) the LLM-based selection of the best answers provided by male and female users for a given question.
For this purpose, we employ Amazon Mechanical Turk (MTurk), a well-known crowdsourcing platform where tasks are delegated to a distributed workforce who complete them remotely for payment. 
To ensure ethical compensation, we adhered to a minimum wage equivalent to at least 8\$ per hour for participants.

Each task in our experiments consists of ten questions, including one \emph{attention-checking} question designed to ensure that participants carefully read and understood the task instructions. 
The use of attention-checking questions is essential for maintaining the reliability of the evaluations, as MTurk workers are typically incentivized to complete tasks quickly since they are paid per completed task, not by the hour \cite{gibson2011using}. This creates a natural tendency to prioritize speed over thoroughness. For each task, we exclude from the analysis the assessments of workers who either failed the attention-check question or consistently selected the same answer.
Moreover, to enhance the robustness of our analysis, we selected \emph{Masters} workers who have consistently demonstrated high accuracy and reliability across various tasks.

Finally, for both tasks, we assess inter-rater reliability between human and non-human annotators using Krippendorff’s alpha~\cite{krippendorff2004reliability}, which is well-suited for varying numbers of raters and missing data.

\subsection{Gender Inference Assessment}
\label{subsec:mt_gender}

In this section, we present the human evaluation conducted to assess the effectiveness of \GC in gender inference. While it is a widely used tool, this evaluation provides a more detailed analysis of its performance, allowing for the quantification of errors across different genders.

We selected 440 user names, including 44 widely recognizable attention-checking names (such as ``John = male''), with a distribution of 80\% male, 10\% female, and 10\% unknown, following the proportion of the entire dataset. The names were divided into 44 tasks, each containing ten names, including one attention-checking name.
Each task was completed by five workers, who classified names as  ``Male'',  ``Female'' or ``Neither'' and rated their confidence as ``Very'', ``Fairly'', or ``Not confident''.
To assign a final gender to a name, we compute scores based on confidence: 2 for ``Very'', 1 for ``Fairly'', and 0 for ``Not Confident''. 
The gender with the highest score was taken as the final assignment, while tied or nearly tied scores (i.e., one point difference) were categorized as ``Neither'' to reflect a lack of consensus.

\begin{table}[]
\centering
\caption{Accuracy (Acc.), Precision (P), and Recall (R) of \GC in comparison to human evaluation for gender inference. Class-level metrics also include the F1-score (F1). The \emph{Corrected Overall} row shows updated metrics for \GC after manual corrections to human evaluation concerning the Female class.}
\label{tab:gender_human}
\begin{tabular}{lrrrr}
\toprule
 & Acc. (\%) & P (\%) & R (\%) & F1 (\%) \\ \midrule
Overall           & $84$ & $70$& $70$ & - \\ 
\cdashlinelr{1-5}
\hspace{0.2cm} ``Male''        & -                        & $93$ & $90$ & $92$                  \\
\hspace{0.2cm} ``Female''      & -                        & $79$ & $59$ & $68$                  \\ 
\hspace{0.2cm} ``Neither''     & -                        & $39$ & $61$ & $48$                  \\ \midrule 
Corrected Overall & $86$ & $77$ & $74$ & -  \\
\cdashlinelr{1-5}

\hspace{0.2cm} ``Male''        & -                        & $90$ & $94$ & $92$  \\
\hspace{0.2cm} ``Female''      & -                        & $80$ & $83$ & $81$  \\
\hspace{0.2cm} ``Neither''     & -                        & $61$ & $44$ & $51$                  \\
\bottomrule
\end{tabular}
\end{table}

Table \ref{tab:gender_human} highlights \GC's strong overall performance with 84\% accuracy. However, \GC shows notable disparities between genders: the male class achieves an F1-score of 92\%, while the female class achieves 68\%, reflecting challenges in accurately classifying female names.

A manual analysis of cases where humans disagreed with \GC’s predictions for females (18 out of 44 names) revealed that \GC correctly inferred the gender in 9 instances. We corrected the human assessment of these instances, increasing the overall accuracy to 86\%, the female Recall from 59\% to 83\%, and the F1-score to 81\%. 

The ``Neither'' class remains challenging, with an F1-score of 48\% initially, rising to 51\% after corrections, indicating difficulties in handling ambiguous names. Interestingly, human annotators tend to assume men as a default~\cite{perez2019invisible}, which can be seen in their categorization of ambiguous or fantasy names, contrasting with \GC, which relies on statistical matching. In fact, in 17 cases where \GC could not assign a gender, humans labeled 14 names as Male and 3 as Female.
For the purposes of this study, however, only male and female classifications are considered.

Finally, we computed Krippendorff’s alpha to assess inter-rater agreement among both human and non-human annotators. We included \llama and \mistral as additional annotators for the binary gender inference task, applying the same procedure as above. \llama achieved an accuracy of 88\%, with an overall precision and recall of 78\%. It performed well across genders, yielding F1-scores of 94\% for males and 78\% for females. By contrast, \mistral reached 80\% accuracy, with a precision of 77\% and a recall of 66\%. However, it showed a markedly lower recall for females (33\%, F1-score 49\%), reflecting a tendency to under-identify female names by often classifying them as \textit{Neither}.

Among human raters, we observed a moderate agreement ($\alpha=$ 0.52). Incorporating the two LLMs slightly reduced the agreement to 0.50, suggesting that LLMs perform similarly to humans on this relatively straightforward task.

In conclusion, the human evaluation supports the accuracy and reliability of \GC in inferring gender.

\subsection{Answer Quality Assessment}
\label{susec:mt_ranking}

The purpose of the human evaluation of answer quality in relation to their corresponding questions is twofold. First, unlike \textsc{Stack Overflow}, human evaluators are presented only with the textual content of the questions and answers. This removes users' metadata information, which can introduce biases. By eliminating these factors, the evaluation aims to determine whether \textsc{Stack Overflow} users have occasionally selected an answer as the “best” due to unrelated factors, even when a better alternative exists.

Second, human evaluation serves as a benchmark for LLMs' ability to perform the answer quality assessment. LLMs have demonstrated strong reasoning skills, particularly in complex problem-solving domains like programming. By comparing human judgments with LLM-generated assessments on a curated subset of questions, the evaluation can help validate LLMs' reliability in this task. Given the resource-intensive nature of large-scale human evaluation, this benchmarking step is essential for enabling scalable and trustworthy analysis across the entire dataset.

We analyzed 220 questions, including 22 attention-checking questions, to assess MTurk worker reliability. Questions were selected from the \accdate dataset, assuming the asker accepted the answer immediately upon posting. To simplify the task, we included only questions with two answers and limited the text length to $1,000$ characters for both questions and answers. Questions containing direct links to \textsc{Stack Overflow} were excluded to prevent evaluators from identifying the source and ensure unbiased assessments. Attention-checking questions were chosen where all LLM models and \textsc{Stack Overflow} agreed on the best answer. Finally, we maintained a balanced gender distribution, selecting 110 questions with female-provided accepted answers and 110 with male ones.

We targeted workers with information technology-related jobs to align with the analysis's specific domain (i.e., coding questions). 
As done for the gender inference assessment, each task consisted of ten questions, each assessed by ten workers. Users were asked to select the best answer for each question and rate their confidence as ``Very'', ``Fairly'', or ``Not''. The final assessment was assigned to the answer with the highest cumulative confidence score.

We excluded all workers who failed the attention check or always selected the same answer, resulting in a final set of 184 evaluated questions. For each LLM-based model, we included only the questions where the model consistently selected the same best answer across three runs with shuffled answer orderings (as described in Section~\ref{sec:methodology}). This resulted in varying numbers of evaluated questions across models, as shown in Table~\ref{tab:ranking_human_ass}.

\begin{table}[]
\centering
\caption{Accuracy (Acc), Precision (P), Recall (R), and number of questions successfully elaborated (\#Q) of each Source/Model when compared to human assessment of the best answer.}
\label{tab:ranking_human_ass}
\begin{tabular}{lrrrr} \toprule
Source/Model & Acc. (\%) & P (\%) & R (\%) & \#Q\\ \midrule
\textsc{Stack Overflow}    & 66     & 67      & 66   &  184 \\
\monotf       & 54     & 54      & 55   &  183 \\
\rankzep      & 59     & 59      & 59   &  130 \\
\rankvic      & 62     & 62      & 62   &  146 \\
\llama        & 76     & 76      & 76   &  104 \\
\mistral      & 68     & 68      & 69   &  127 \\ \bottomrule
\end{tabular}
\end{table}

Human evaluations serve as the ground truth to assess the performance of various sources (e.g., \textsc{Stack Overflow}) and methods (e.g., LLM-based models) for selecting the best answers. Table \ref{tab:ranking_human_ass} presents the Accuracy (Acc.), Precision (P), and Recall (R) for each ranking source, expressed as percentages, together with the number of questions (\#Q) each model successfully elaborated.

\textsc{Stack Overflow} achieves an accuracy of 66\%, indicating that community-selected ``best answers'' align with human evaluations about two-thirds of the time. While this suggests a reasonably reliable process, the remaining 34\% of disagreements deserve further investigation, particularly to explore potential biases arising from the community’s exposure to answerer metadata.

Among the LLM-based models, \llama achieves the highest accuracy at 76\%, followed by \mistral with 68\%. \rankvic and \rankzep, both based on LLaMA-v2 \cite{touvron2023llama} and fine-tuned for ranking, follow with accuracies of 62\% and 59\%, respectively. \monotf, a traditional document ranking model, performs the worst with an accuracy of 54\%, likely due to its limited suitability for technical CQA answer evaluation.
Precision and Recall across all sources remain stable, indicating their ability to identify relevant answers without disproportionately including incorrect ones. 

To further analyze model behavior, we examined performance by gender using the F1-score. Since each question includes at least one male and one female-authored answer, disagreement with human labels allows us to see whether a model tends to favor one gender over the other. \llama demonstrates balanced performance, with F1-scores of 77\% for female-authored and 75\% for male-authored answers. \mistral shows similar behavior: 69\% for females, 68\% for males. In contrast, \textsc{Stack Overflow} shows an opposite trend with 67\% for males and 66\% for females. \rankvic and \rankzep show relatively balanced results across genders, while \monotf performs the worst, especially on female-authored answers (53\% F1).
Overall, models like \llama and \mistral exhibit consistent performance across all metrics, closely matching human evaluations. This consistency suggests that their assessments remain stable and reliable when scaled to larger datasets. 

Finally, the inherently subjective nature of this task is reflected in the inter-rater agreement. While the gender classification task yielded moderate agreement (Krippendorff's $\alpha=$ 0.52 among humans), the answer quality task achieved a much lower $\alpha=$ 0.30, dropping to $\alpha=$ 0.26 when including LLMs. This confirms that selecting the best answer is inherently ambiguous, often involving multiple valid options and subjective preferences.
\section{Experimental Results}
\label{sec:results}

In this section, we present the results of the \emph{Features-} and \emph{LLMs-}based analysis.

\subsection{Features-based Analysis Results}
\label{subsec:features_analysis_res}

\begin{table*}[!htbp]
\centering
\caption{Results of \emph{Mann–Whitney U} test (5,000 permutations and Bonferroni correction) comparing features between genders on two different SO-based datasets. For each feature and each gender, we report the mean value, the difference in percentage (``\%Diff''), and the $p$-value. For ``\%Diff'', a ``+'' indicates an advantage for males, and a ``-'' for females.}
\label{tab:ttest}
\resizebox{\textwidth}{!}{%
\begin{tabular}{lrrrrrrrr}
\toprule
\multicolumn{1}{c}{\textbf{2017-2024}} & \multicolumn{1}{c}{\textbf{SO-Reputation}} & \multicolumn{1}{c}{\textbf{Views}}                      & \multicolumn{1}{c}{\textbf{UpVotes}}              & \multicolumn{1}{c}{\textbf{DownVotes}}                 & \multicolumn{1}{c}{\textbf{Answers}} & \multicolumn{1}{c}{\textbf{AcceptedAns}}          & \multicolumn{1}{c}{\textbf{AvgScore}} & \multicolumn{1}{c}{\textbf{AvgDelay}}     
\\ 

\midrule Male & \underline{1,529.21} & \underline{186.43} & \underline{145.36} & \underline{20.57} & \underline{29.64} & \underline{10.51} & \underline{2.04} & \underline{6,924.80} \\
Female  & 972.99 & 145.90 & 110.21 & 18.44 & 20.18 & 6.74 & 1.69 & 6,723.13 \\
\%Diff & {+36.37} & {+21.74} & {+24.18} & {+10.35} & {+31.88} & {+35.87} & {+17.16} & {+2.91} \\
p-value & $<0.01$ & $<0.01$ & $<0.01$ & $<0.01$ & $<0.01$ & $<0.01$ & $<0.01$ & $<0.01$ \\ \midrule

\multicolumn{1}{c}{\textbf{AccDate}}   & \multicolumn{1}{c}{\textbf{SO-Reputation}} & \multicolumn{1}{c}{\textbf{Views}}                      & \multicolumn{1}{c}{\textbf{UpVotes}}              & \multicolumn{1}{c}{\textbf{DownVotes}}                 & \multicolumn{1}{c}{\textbf{Answers}} & \multicolumn{1}{c}{\textbf{AcceptedAns}}          & \multicolumn{1}{c}{\textbf{AvgScore}} & \multicolumn{1}{c}{\textbf{AvgDelay}}                      \\ \midrule

Male& \underline{10,410.07}& \underline{1,492.72} & \underline{604.61} & \underline{243.01} & \underline{261.71} & \underline{114.91} & \underline{2.13} &{2,725.58}  \\
Female & 3,068.12 & 429.19 & 290.40 & 95.80 & 76.99 & 29.96 & 1.92 & \underline{3,678.96} \\
\%Diff & +70.53 & +71.25 & +51.97 & +60.58 & +70.58 & +73.93 & +9.86 & -34.98  \\

p-value & $<0.01$& $<0.01$ & $<0.01$& $<0.01$ & $<0.01$ & $<0.01$ & $<0.01$ & $0.34$ \\               \bottomrule
\end{tabular}
}
\end{table*}

We report the results of our features-based analysis (Section~\ref{subsec:features_analysis}) in Table~\ref{tab:ttest}. For two datasets, namely \texttt{2017-2024} and \accdate, we present the mean value of each considered feature for male and female users, together with their percentage difference (\%Diff), where a “+” indicates an advantage for males, and a “–” for females. Table~\ref{tab:ttest} also reports the $p$-value derived from the \emph{Mann–Whitney U} test, conducted at commonly applied significance levels of \(\alpha\). To enhance readability, we omit the statistics for the \accday dataset, which are identical to those of \accdate.

Over the \texttt{2017-2024} period, we find statistically significant differences between male and female users across all features, with male users consistently exhibiting higher mean values. For instance, male users have on average $36\%$ higher reputation and $35\%$ more accepted answers compared to female users.
In the \accdate dataset, the same trends emerge, except for the AvgDelay feature, which is higher for female users. Since all features are computed considering each user’s entire activity history (\texttt{2008-2024}), Table~\ref{tab:ttest} also highlights that the \accdate dataset captures the most active users (i.e., higher reputation, visibility, and volume of contributions, both in the number of answers provided and in response times).
Furthermore, our statistical analysis of question-level features (\emph{Score}, \emph{Views}, and number of \emph{Answers}) reveals that the sets of questions included in \accdate and \accday have significantly higher mean values across all dimensions, indicating that these subsamples also concentrate on more popular questions.

Moreover, unlike the results reported in Table~\ref{tab:ttest}, if we recompute features such as the number of Answers and Accepted Answers, AvgScore, and AvgDelay using only the data from the \accdate and \accday subsamples, the picture changes. In these cases, female users display higher activity levels, with \emph{\%Diff} values of $+83.98$ and $+94.42$ for the number of provided answers and $+121.43$ and $+100$ for the number of accepted answers, respectively.  
This phenomenon can be explained by two factors. First, as shown in Table~\ref{tab:dataset_statistics}, restricting the dataset to questions with both male and female respondents increases the proportion of female users to about 30\%. Filtering in this way alters the sample composition, disproportionately affecting the majority group (males) and effectively concentrating the analysis on questions where females are already active. Second, when we replicated this analysis on questions from the \texttt{2008-2024} and \texttt{2017-2024} periods but restricted to those involving both male and female respondents, we observed the same pattern: female users consistently showed higher activity and contributed more accepted answers. 

It is important to emphasize that this filtering process does not introduce bias into our analysis but is instead a necessary methodological step. Our central research question is to investigate whether, when both a male and a female answer are available for the same question, the community systematically favors one gender over the other in the selection of the accepted answer. To address this, we must restrict our dataset to cases where at least one male and one female response co-occur in the same thread. Without such filtering, the analysis would be dominated by questions answered exclusively by male users, reflecting the overall gender imbalance on the platform rather than the decision-making process of askers when presented with alternatives.  Thus, the apparent increase in female activity and recognition in the \accdate and \accday datasets is not an artifact of bias but a direct consequence of focusing on the subset of questions that have answers written by men and women.

\paragraph{Homophily.} The behavior emerging in \accdate aligns with prior  findings~\cite{morgan2017programming,ford2017someone}, suggesting homophilic tendencies among women: they are more likely to answer questions where other female users are involved, thereby amplifying their presence in these question threads. 
To assess gender homophily in answering behavior, we first computed two standard Social Network Analysis metrics: attribute assortativity and Spearman rank correlation on answerer-asker gender pairs. 

Results for both datasets (\accdate and \accday) show negligible values (assortativity$<0.01$, $\rho<0.01$) and non-significant correlations ($p>0.67$), indicating no detectable gender-based assortative mixing in the directed interaction network.
However, we note that these measures are normalized relative to a random mixing baseline, and may fail to capture homophilic patterns in cases of strong gender imbalance such as ours. In fact, while the proportion of answers is relatively balanced among genders (53.78\% of answers by males in the \accdate dataset), the distribution of questions is highly skewed, with male askers accounting for approximately 80\% of questions in both datasets. This imbalance 
highlights the need to carefully account for the baseline expectation when analyzing homophilic tendencies in answering behavior.

For this reason, we further investigate homophilic tendencies by computing the probability of users answering questions posed by male or female askers.
To assess same-gender answering preference, we calculate the expected behavior, assuming users answer questions without regard to the gender of the asker. For each user $u$, we compute the \textit{User Answering Ratio} (UAR), which is defined as
$$
\text{UAR} (u) = \frac{|A^u_h|}{|A^u|}
$$
where $A^u$ is the set of answers provided by $u$, while $A^u_h$ is the set of homophilic answers for $u$, i.e., where the gender of the asker corresponds to the gender of $u$.
With this premise, we define the \textit{Answering Ratio} for a gender $g$ as the mean of the UAR  of users with gender $g$, formally defined as
\begin{equation}
\label{eq:a_ratio}
    \text{Answering Ratio} (g) = \frac{1}{n_g}\sum_{i=1}^n \mathds{1} (\text{gender}(u_i) = g) \cdot \text{UAR}(u_i) 
\end{equation}
where $\mathds{1} (\text{gender}(u_i) = g)$ is $1$ if the gender of the $i$-th user is $g$ and $0$ otherwise, $n_g$ is the number of users with gender $g$, and $n$ is the total number of users.
Similarly, we define 
\begin{equation}
\label{eq:q_ratio}
    \text{Question Ratio(g)} = \frac{|Q^g|}{|Q|}
\end{equation}
which represents the proportion of questions asked by users of a specific gender \(g\) in the dataset. 
If answering behavior is gender-unaware, the $\text{Answering Ratio(g)}$ should equal the $\text{Question Ratio(g)}$. This equality serves as the baseline for gender-unaware answering behavior.
Accordingly, to quantify homophilic tendencies, we define a \textit{``same-gender answer preference''} metric $P_{g}$ computed as the ratio between Eq. \ref{eq:a_ratio} and Eq. \ref{eq:q_ratio}.
A value of \(P_g > 1\) indicates a preference for answering questions from individuals of the same gender, while a value less than 1 suggests a preference for the opposite gender.

The results reveal that males exhibit balanced answering behavior, with $P_m =$ 1.01 in both datasets. In contrast, for female users, homophily emerges: they are more likely to answer female askers, with a $P_w =$ 1.19 for the \accdate dataset and $P_w =$ 1.32 for the \accday dataset.
These results indicate that the observed higher activity and recognition levels among female users are driven by both the dataset's compositional effects and a pronounced homophilic tendency, amplifying their visibility and influence within specific subsets of interactions.

Finally, even within these subsets, males still achieve higher average scores, supporting the hypothesis proposed by \cite{brooke2021trouble}, indicating a subtle bias in scoring that disadvantages female contributors.

\subsection{LLM-based Analysis Results}

Table \ref{tab:accdate_res} presents the results for the \accdate and \accday datasets. 
The \#Questions column reports the number of questions the models successfully elaborated. All models, except \monotf, failed to elaborate on some questions, which reduced the size of the datasets. In our experiments, error rates ranged from $1.41\%$ to $18.34\%$. 
After the preprocessing (explained in Section~\ref{sec:methodology}) in \accdate, the percentage of discarded questions ranged from 25.92\% (\rankzep) to 58.34\% (\mistral). Similarly, in \accday, the loss ranged from 27.14\% (\rankzep) to 61.76\% (\mistral). While this filtering step removes a non-negligible share of data, we believe it ensures the most robust and interpretable foundation for the analysis. Importantly, we verified that loosening or tightening the preprocessing criteria leads to comparable final results and conclusions, reinforcing the stability of our findings.

The models agree with \textsc{Stack Overflow}’s best answer (P@1 SO-Match column) in 48.80\% to 68.16\% of questions for the \accdate dataset and in 47.58\% to 63.65\% for the \accday dataset.
In cases of model-SO mismatch, we analyzed whether a female user wrote the accepted answer with the model preferring a male-authored response (Top Ranked Male - Accepted Female column) or vice versa (Top Ranked Female - Accepted Male column). 
Percentages for both configurations are consistently close across models and datasets, with differences in percentage ($\Delta$ column) ranging from 0.08\% to 1.55\% for the \texttt{AccDate} dataset and 0.02\% to 1.69\% for the \texttt{Acc1Day} dataset.

\begin{table*}[!ht]
\centering
\caption{LLM-based analysis results for each dataset and model. The column \#Questions indicates the number of questions correctly elaborated by the model. The \textit{P@1 (SO Match)} column reports the percentage of questions where the models agree with \textsc{Stack Overflow} best answer. The \textit{Top Ranked Male - Accepted Female} and \textit{Top Ranked Female - Accepted Male} columns report the percentages of cases where the model's top-ranked answer differs from \textsc{Stack Overflow}'s accepted answer based on the answerer's gender. The \(\Delta\) column shows the percentage difference between the two cases.}
\label{tab:accdate_res}
\begin{tabular}{llccccc}
\toprule
Dataset & Model & \#Questions & \thead{P@1\\(SO Match)} & \thead{Top Ranked Male  \\ Accepted Female (\%)} & \thead{Top Ranked Female \\ Accepted Male (\%)} & $\Delta$ (\%)  \\ \midrule
\multirow{5}{*}{AccDate} 
& \monotf               & 16,520 & 48.80 & 23.92               & \underline{24.00}    &  0.08\\
& \rankvic              & 13,839 & 56.26 & \underline{20.83}   &  20.27               &  0.56\\
& \rankzep              & 18,146 & 54.73 & \underline{21.42}   &  21.01               &  0.41\\
& \llama                & 14,355 & 68.16 & \underline{15.63}   &  14.32               &  1.31\\
& \mistral              & 10,204 & 62.21 & \underline{19.25}   &  17.70               &  1.55\\ 
\midrule
\multirow{5}{*}{Acc1Day} 
& \monotf               & 32,804 & 47.58 & 22.51 & \underline{24.20} & 1.69 \\
& \rankvic              & 26,431 & 55.40 & 19.74 & \underline{20.71} & 0.97 \\
& \rankzep              & 34,937 & 54.72 & 20.02 & \underline{20.81} & 0.79 \\
& \llama                & 26,679 & 63.65 & 16.58 & \underline{16.60} & 0.02 \\
& \mistral              & 18,357 & 59.88 & 19.37 & \underline{19.51} & 0.14 \\
\bottomrule
 \end{tabular}
\end{table*}

In the \accdate dataset, most models more frequently rank male answers above female-accepted ones, with the exception of \monotf, which shows a small opposite effect. In contrast, the \accday dataset shows a consistent trend across all models, where LLMs more often rank female answers above male-accepted ones.
Given that the \accday dataset includes answers posted within a day of the accepted answer, this could suggest that female users often provide high-quality answers after initial responses from male users.
This behavior is supported by our analysis: in the \texttt{2008-2024} and \texttt{2017-2024} periods, male users posted the first answer in over 91\% of cases. Even in the more balanced subsets used in \accdate and \accday, where each thread includes at least one male and one female answer, we still observe that male users were first to answer in approximately 53\% and 55\% of cases, respectively.
With \textsc{Stack Overflow}’s emphasis on quick responses, females' later answers might be less likely to be accepted.

Overall, the minimal $\Delta$ values suggest that \textsc{Stack Overflow} users prioritize solution quality over the answerer’s gender, reinforcing the platform’s meritocratic nature. Instead, observed discrepancies in recognition, such as SO-Reputation, are likely due to differences in response timing and activity levels rather than inherent gender bias.

As an additional check for potential bias in answer selection, we analyzed whether earlier answers were overlooked in favor of later ones from a different gender. Since we lack the exact timestamp of acceptance, we approximated it using the timestamp of the accepted answer. In the \texttt{2017-2024} sample, both male and female accepted answers had an average delay of approximately five days, showing no timing advantage by gender. We then examined cases where an earlier answer was ignored in favor of a later accepted one. In 8.1\% of cases, a female answer was accepted after a male had already answered; in 7.9\%, a male answer was accepted after a female. These near-identical values, combined with the small and mixed $\Delta$ values reported earlier, do not indicate a systematic bias in favor of one gender over the other in the community's acceptance behavior.

Moreover, in addition to the knowledge-sharing nature of \textsc{Stack Overflow}, users are often motivated by self-serving goals, such as providing high-quality answers to enhance their reputation within the community. As a result, answers to the same question are frequently of high quality. In such cases, the decisive factors in selecting the best answer may be subjective.
To investigate this, we analyzed whether the selection of the best answer was correlated with answer length or the SO-Reputation of the responder (which is visible to the asker). For the \accdate dataset, 63\% of the time, the accepted answer was the longest, averaging 1160 characters compared to 754 for non-accepted answers. Similarly, in 56\% of cases, the top responder had a higher SO-Reputation, with an average  of 71,137 compared to 44,485 for others. Similar trends were observed in the \accday dataset.
While answer length plays a stronger role, reliance on SO-Reputation may introduce unintended bias, as male users generally have higher reputation scores. Although reputation can serve as an indicator of reliability, it risks reinforcing gender disparities. 

An important consideration is that some or all of the models used in this study may have been trained on \textsc{Stack Overflow} data, given its availability as a public dataset. This overlap raises the possibility that the models might indirectly know the ground truth answers from \textsc{Stack Overflow}, potentially influencing their ability to rank answers in a way that aligns with \textsc{Stack Overflow}’s accepted answers.
However, as demonstrated by the results in Table \ref{tab:accdate_res}, the models do not simply replicate the accepted answers from \textsc{Stack Overflow}.
Instead, they exhibit a notable degree of generalization beyond the potentially observed data, as evidenced by the cases where model rankings diverge from the answers chosen by the \textsc{Stack Overflow} community. This divergence suggests that while the \textsc{Stack Overflow} data may inform models, models perform their own relevance evaluation that can lead them to alternative rankings, often counteracting the accepted answer in cases where they consider different responses more relevant or informative.
In conclusion, human selections and LLM rankings suggest that answers from men and women are comparable in quality. This finding contrasts with \textsc{Stack Overflow}’s scoring system, which favors men due to higher activity levels and contribution quantity rather than quality. While the community and LLMs prioritize merit, the reputation system inadvertently emphasizes quantity, contributing to gender disparities in recognition.

\section{Conclusion and Discussion}
\label{sec:conclusion}

In this study, we advanced the analysis of gender disparities on \textsc{Stack Overflow} by comparing the quality of answers provided by male and female users to the same questions. We assessed answer quality using state-of-the-art LLMs across questions with male and female respondents.
We conducted two human evaluations to validate key aspects of this study: (i) the inference of users' gender and (ii) the use of state-of-the-art LLMs as a proxy for human evaluation in assessing answer quality. 
The human assessment supports the accuracy of the \GC tool in inferring binary gender, achieving 84\% overall accuracy, with F1 scores of 92\% for male names and 81\% for female names. Additionally, both \textsc{Stack Overflow} and LLMs demonstrated reasonable alignment with human evaluations, with accuracy ranging from 54\% for \monotf to a maximum of 76\% for \llama.
Next, we performed a statistical analysis showing that male users generally have higher values in features related to recognition of their contributions, greater activity levels, and shorter delays in answering. Moreover, the LLM-based analysis indicates that LLM-based models align with \textsc{Stack Overflow}’s accepted answers for approximately 57\% of the questions on average, reaching up to about 68\% in some cases. In instances where model rankings differ from \textsc{Stack Overflow}’s choice, the proportion of cases favoring male versus female answers is balanced, with a maximum discrepancy of only 1.69\%. These results suggest that \textsc{Stack Overflow} users seem to evaluate answer quality independently of the perceived gender of the answerer. Additionally, the findings suggest that a combination of subjective and objective factors (i.e., clarity of explanation and correctness) influences the selection of the best answer. In many cases, questions have multiple answers that are highly relevant, and our experiments do not show any specific criterion, from the ones examined, that determines their choice in these cases.
However, timing plays a significant role, as high-quality responses posted after another answer has been accepted may be overlooked.

In conclusion, we show that both men and women equally provide high-quality answers. The observed bias in recognition metrics is likely tied to \textsc{Stack Overflow}’s reputation system, which emphasizes the quantity of activity, favoring more active users, typically men. 
Moreover, as reported in Section~\ref{sec:relatedwork}, \textsc{Stack Overflow} is often perceived as an elitist, reputation-based platform that primarily, and possibly by design, rewards high-volume contributors. This underscores the need for more refined reputation criteria that move beyond simple activity metrics. Currently, reputation is awarded mainly through voting activity (upvotes on answers or questions +10 points) and accepted answers (+15 points)~\footnote{\url{https://stackoverflow.com/help/whats-reputation}}, a mechanism that tends to favor users with higher posting volumes. We argue that decoupling activity volume from reputation scores and adopting a multi-dimensional descriptor of user roles would better capture the diversity of contributions to the community and prevent scores from indirectly favoring men over women. For instance, welcoming newcomers or mentoring users on how to ask and answer effectively could be recognized as valuable non-technical contributions. In the long term, such an approach would foster a healthier and more inclusive environment, while also benefiting the platform commercially, as perceived elitism may discourage participation from new and female users.

\smallskip\noindent\textbf{Limitations and Future Works.}
Our analysis focuses exclusively on \textsc{Stack Overflow}, the largest CQA platform and the one most commonly studied in prior research. While its scale and relevance make it a strong case study, we acknowledge that platform-specific dynamics may limit the generalizability of our findings. Extending this analysis to other CQA platforms represents an important direction for future work. Moreover, a central limitation concerns gender inference. Because \textsc{Stack Overflow} does not provide gender information, our study relies on name- and country-based inference tools to estimate gender. This approach yields a binary categorization (male/female), which does not capture the full spectrum of gender identities.
We recognize that gender is a complex and socially constructed concept that cannot be reduced to name-based assumptions. Moreover, users may intentionally obscure their gender or adopt pseudonyms that do not correspond to real names or binary categories.
However, our intent is not to assign or assume users’ actual gender identities, but rather to model the likely perception of gender by other users. Since community behavior is shaped by such perceived signals (e.g., usernames), analyzing the effects of inferred gender remains meaningful for studying bias in recognition.

In future work, we aim to address these limitations by using other methodologies, possibly qualitative, to study the experience of various gender identities. This is fundamentally different from the research on this paper, which is about potential biases stemming from gender perceptions.
Moreover, future work could explore two promising directions: (i) extending the Social Network Analysis to better understand community dynamics, e.g., by employing Exponential Random Graph Models~\cite{lusher2013exponential} to statistically validate our findings on gender-related homophily; and (ii) conducting user studies to investigate which factors influence the selection of the best answer.

\smallskip
\noindent\textbf{Code and data availability.}
All code used in this study\footnote{\url{https://github.com/maddalena-amendola/Gender-Disparities-StackOverflow}} as well as the results of the MTurk assessments will be released with the camera-ready version to ensure full reproducibility. To protect user privacy, gender inference data will be shared only in an aggregate and anonymized form, without linking usernames to inferred genders.

\smallskip
\noindent\textbf{Hardware and compute time.}
Our experiments ran on a server with two Intel Xeon Platinum 8480CL CPUs and an Nvidia H100 GPU. Most of the compute time was spent on LLMs (\llama and \mistral), taking about one day on \accday and half a day on \accdate.

\paragraph{\textbf{Licensing and Attribution.}} We utilized content from \textsc{Stack Overflow}, licensed under CC BY-SA 4.0. The models employed in our research include \monotf (Apache 2.0), \rankvic (LLaMa 2 CL), \rankzep (MIT), \llama (Meta LLaMa 3 CL), and \mistral (Apache 2.0). All resources were used in compliance with their respective licenses.
\appendix
\section*{Appendix A: Posts and Users Data}

This study relies on the \texttt{Posts} and \texttt{Users} tables from the Stack Exchange public data dump\footnote{\url{https://meta.stackexchange.com/questions/2677/database-schema-documentation-for-the-public-data-dump-and-sede}}.
Posts are divided into \emph{Questions} and \emph{Answers} according to the field \texttt{PostTypeId}.

\subsection*{Posts}

\textbf{Common fields (both Questions and Answers):}
\begin{itemize}
  \item \texttt{Id}: unique identifier of the post;
  \item \texttt{PostTypeId}: distinguishes the post type (\texttt{1}=Question, \texttt{2}=Answer);
  \item \texttt{CreationDate}: timestamp when the post was created;
  \item \texttt{Score}: net score (upvotes minus downvotes);
  \item \texttt{Body}: full content of the post in HTML;
  \item \texttt{OwnerUserId}: identifier of the user who created the post (NULL if deleted);
  \item \texttt{OwnerDisplayName}: fallback display name when the owner is deleted;
  \item \texttt{LastEditorUserId}: identifier of the most recent editor; 
  \item \texttt{LastEditDate}: timestamp of the last edit;
  \item \texttt{LastActivityDate}: last time the post was active (e.g., comment, edit, new answer);
  \item \texttt{CommentCount}: number of comments associated with the post;
  \item \texttt{ContentLicense}: license under which the content is released;
  \item \texttt{DeletionDate}: date when the post was deleted (only available in \texttt{PostsWithDeleted}).\\
\end{itemize}

\textbf{Question-specific fields (\texttt{PostTypeId=1}):}
\begin{itemize}
  \item \texttt{AcceptedAnswerId}: identifier of the answer accepted by the author;
  \item \texttt{ViewCount}: number of times the question has been viewed;
  \item \texttt{Title}: title of the question;
  \item \texttt{Tags}: tags assigned to the question;
  \item \texttt{AnswerCount}: number of (non-deleted) answers;
  \item \texttt{FavoriteCount}: number of times the question was favorited/bookmarked;
  \item \texttt{ClosedDate}: date when the question was closed (if applicable);
  \item \texttt{CommunityOwnedDate}: date when the post became community wiki (if applicable).\\
\end{itemize}

\textbf{Answer-specific fields (\texttt{PostTypeId=2}):}
\begin{itemize}
  \item \texttt{ParentId}: identifier of the corresponding Question.
\end{itemize}

\subsection*{Users}

The \texttt{Users} table contains metadata for each account:
\begin{itemize}
  \item \texttt{Id}: unique identifier of the user;
  \item \texttt{Reputation}: total reputation score of the user;
  \item \texttt{CreationDate}: date when the user account was created;
  \item \texttt{DisplayName}: chosen username displayed publicly;
  \item \texttt{LastAccessDate}: timestamp of the most recent user activity;
  \item \texttt{WebsiteUrl}: user-provided website link;
  \item \texttt{Location}: self-declared location;
  \item \texttt{AboutMe}: user-written biography;
  \item \texttt{Views}: number of times the user profile was viewed;
  \item \texttt{UpVotes}: total number of upvotes cast by the user;
  \item \texttt{DownVotes}: total number of downvotes cast by the user;
  \item \texttt{ProfileImageUrl}: link to the user’s avatar image;
  \item \texttt{AccountId}: network-wide identifier across Stack Exchange sites.
\end{itemize}

\vspace{0.5em}
Posts are linked to Users via the field \texttt{OwnerUserId}.

\bibliographystyle{abbrv}
\bibliography{biblio}

\end{document}